# Coherence loss of partially mode-locked fibre laser

Lei Gao[1], Tao Zhu[1*], Stefan Wabnitz[2], Min Liu[1], and Wei Huang[1]

[1]Key Laboratory of Optoelectronic Technology & Systems (Ministry of Education), Chongqing University, Chongqing 400044, China.

[2]Dipartimento di Ingegneria dell'Informazione, Università degli Studi di Brescia and INO-CNR, via Branze 38, 25123 Brescia, Italy.

Stochastically driven nonlinear processes limit the number of amplified modes in a natural system due to competitive mode interaction, which is accompanied by loss of coherence when increasing the complexity of system. Specifically, we find that modulation instability, which exhibits great fluctuations when it is spontaneously grown from noise in conservative systems, may possess a high degree of coherence in dissipative laser system with gain. Nonlinear mode interactions can be competitive or cooperative: adjusting the intracavity polarization state controls the process of loss of coherence. Single-shot spectra reveal that fibre laser redistributes its energy from center wavelength mode to sidebands through parametric instabilities, and subsequently longitudinal modes are populated via cascaded four-wave-mixing. Parametric frequency conversion populates longitudinal modes with a random distribution of position, intensity and polarization, resulting in partially (rather than highly) coherent pulses. These dynamics unveil a new route towards complex pattern formation in nonlinear laser systems.

The transition from highly coherent to weakly coherent states is common in most natural nonlinear systems, such as in optics [1,2], Bose-Einstein condensation [3], semiconductors [4], free-electron lasers [5], and fluids [6]. For specific conservative and dissipative systems, stable coherent solutions with self-sustaining properties can be observed. For example, solitons are formed owing to a balance between nonlinearity and diffraction/dispersion in Hamiltonian systems; dissipative solitons occur under the additional balance of loss and gain [7-9]. Soliton systems possess a high degree of coherence, owing to well-determined phases among their different longitudinal (or frequency) modes. However, spontaneous or noise driven processes are accompanied by energy redistribution into additional modes, leading to a loss of coherence, that may eventually disrupt coherent soliton-like solutions. For example, consider modulation instability (MI), a process that amplifies noise exponentially upon propagation under the combined action of nonlinearity and dispersion/diffraction [10-13]. Due to its stochastic origin, competitive mode interactions enhance fluctuations in single-shot spectra, until complete incoherence is found in both Stokes and anti-Stokes regions [10].

In a dissipative system such as a laser, pattern formation may be initiated by the coherent excitation of MIs. When multi-wave mixing cascades among the MI gain bands, a pulsed laser output with partial coherence is observed. Output from such a partially mode-locked laser (PML), is sometimes referred to as noise-like pulses, since laser pulses *seem* to originate from the amplification of noise: bunches of pulses with irregular varying time duration and intensity are observed [14-19]. The autocorrelation trace of a PML contains a coherence peak of femtosecond duration, sitting on a broad pedestal with a duration ranging from several to hundreds of picoseconds. Moreover, the optical bandwidth of a PML is often comparable to or even larger than the gain bandwidth of the active fibre [15]. As it represents a non-stationary state in a complex dispersive nonlinear system, understanding the underlying physics of PML is scientifically important but technically challenging. Although several experiments have investigated different properties (e.g., stochasticity, optical rogue waves, turbulence, and periodicity) of PMLs, their underlying physics remains controversial [10,16-19]. The main reason for that is the presence of several competing nonlinear processes, which makes it extremely difficult to observe the dynamics of PML formation. Besides, detecting optical spectra at MHz frequencies is still technically challenging, as it requires the application of dispersive Fourier transformation (DFT), a technique which enables transient spectrum detection [10]. Different interpretations of PMLs based on cavity nonlinear transmission and walk-off between two polarizations [14], on the combined effect of soliton collapse and positive cavity feedback [20], and on the oversaturation of the intracavity saturable absorber (SA) [21], have been proposed. Here, we experimentally prove that PML action originates from vector parametric frequency conversion, and that the mechanism of coherence loss is can be controlled by intracavity polarization rotation.

The mechanism of coherence loss in a PML is similar to the generation of chaotic Kerr combs based on parametric frequency conversion within high quality factor (Q) microresonators. In that context, both highly and partially coherent pulses have been produced, based on cascaded four-wave-mixing (FWM) among parametric gain lobes [22]: an universal route was proposed for the generation of microresonator frequency combs [23]. The main principle of cavity parametric frequency conversion is schematically shown in Fig. 1 (a). Primarily, energy transfers from a continuous wave (CW) central mode into parametric or MI gain sidebands: corresponding longitudinal modes are thus populated. MI is a degenerate FWM (DFWM) process, where two pump photons at frequency of $\omega_0$ are annihilated, and two new photons with Stokes or anti-Stokes frequencies ($\omega_{mS}$ and $\omega_{mAS}$) are simultaneously generated. Both energy conservation and phase-matching conditions are satisfied. In microresonators, the frequency spacing (say, $\Delta$) between the center mode and the primary sidebands is much larger than the spacing between longitudinal modes (or cavity free-spectral range, say, $\delta$). Next, cascaded FWM may occur among the pump and different MI gain lobes. When the pump power is high enough, newly generated frequencies merge into a gap-free or continuum comb.

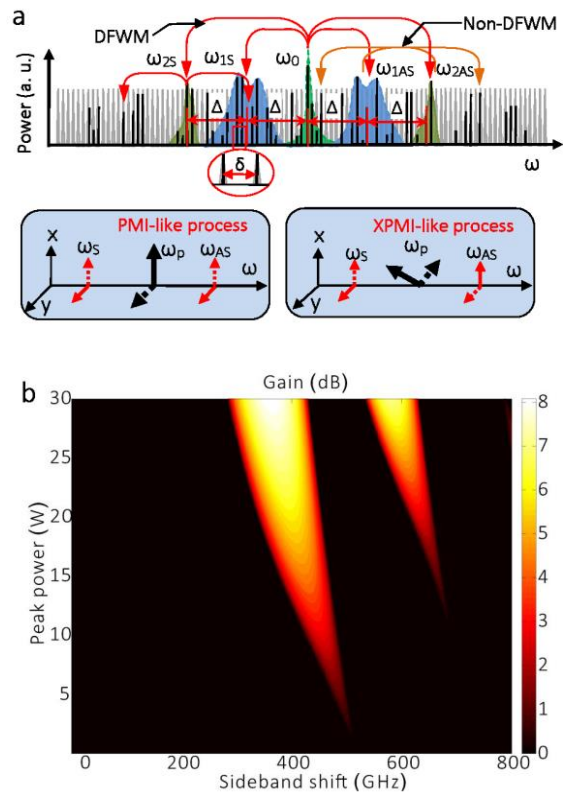

Figure 1. PML dynamics based on stochastic parametric conversion. (a) Primary gain sidebands, $\omega_{1S}$ and $\omega_{1AS}$, are generated from parametric instability: high-order gain lobes are activated with the optimization of the polarization state. Δ would be preserved in all sidebands due to energy conservation. Cascaded FWM populates longitudinal modes between the PI gain lobes. The insets are phase-matching processes of two different vector MI processes. The solid and dotted lines represent two orthogonal polarization components, respectively. In the PMI process, the polarizations of two sidebands are orthogonal to that of the pump, and are parallel to each other. In XPMI, sidebands are polarized at 45° from the pump laser polarization; (b) Numerically computed PI gain sidebands in the laser cavity with periodic dispersion and loss. The calculated sideband shift for the peak power of 5.15 W is about 490 GHz, which is close to the experimental value of 438 GHz. The discrepancy is mainly induced by the estimate error of the average power of pulses.

In time domain, the coherence of the generated pulses depends on both the coherence of parametric gain lobes and the phase-matching condition of cascaded FWM. For microresonator Kerr-combs, adjacent longitudinal modes populated in the MI gain lobes are highly coherent. The coherence of the formed pulses is mainly determined by the appearance of a sequence of sub-combs. When Δ is not an integer multiple of δ, sub-combs with different offset frequencies are generated, resulting in poorly coherent pulses [25]. Highly coherent pulses may be generated within a laser cavity incorporating a high finesse Fabry-Pérot interferometer or a high Q microresonator [24,25]. For a polarization-maintaining fibre laser cavity, scalar cascade FWM occurs. However, both scalar and vector FWM may be supported in non-polarization-maintaining cavity (such as our fibre ring laser), so that both PML and highly coherent pulses may be generated, depending on the orientation of intracavity polarization controllers (PCs). In general, vector FWM processes occur whenever the center mode has an elliptical polarization state. As a matter of fact, two vector FWM processes with cross-coupling between two polarization modes may occur, namely, polarization MI (PMI)-like and cross-phase MI (XPMI)-like, as depicted in the insets of Fig.1a [26]. These processes coexist with scalar FWM and compete with each other, until the winner dominates the whole parametric process. When compared with scalar case, the phase-matching conditions of cascaded PMI-like or XPMI-like FWM are even more complex, and in most cases, partially coherent pulses rather than high coherent pulses are generated due to the loosely fixed phase relation between cavity modes.

In optical fibres with a self-focusing Kerr coefficient, anomalous dispersion is required for the nonlinear phase-matching condition of scalar MI. However, the presence of periodically varying dispersion and loss/gain in a fibre ring resonator can lift this restriction, and a specific kind of MI, parametric instability (PI), appears [27-29]. In this case, the condition for quasi-phase-matching (QPM) of the nonlinear FWM process can be expressed as $\bar{\beta}_2 \Omega_k^2 + 2\gamma P_p = 2\pi k/L$, where, $\bar{\beta}_2$ is the average dispersion, $\Omega_k$ is the pulsation detuning, $\gamma$ is the nonlinear coefficient, $P_p$ is the peak power in the quasi-CW mode, $L$ is the cavity length or dispersion period, and $k$ is the sideband order. Parametric frequency conversion occurs from the central mode into distant cavity modes such that the power-dependent QPM condition is satisfied. Biasing the intracavity polarization controller (PC) varies the nonlinear phase contribution to the QPM condition, since graphene is polarization dependent. Figure 1 (b) illustrates the numerically computed (see Methods section) PI gain within a range of peak powers, with parameters used in this experiment. We find two distinct spectral peaks, which agree quite well with the experimental results.

In this article, we experimentally demonstrate the formation process of PML based on PI and subsequent cascaded vector FWM among the PI gain lobes under cavity detuning control. The detuning can be freely changed by adjusting the state of polarization (SOP) of light passing through the SA, which exhibits a significant nonlinear phase response in our experiment. As we shall see, the dispersion-variation-induced PI that arises in the dissipative laser system exhibits little fluctuations, and quasi-periodic cooperative and competitive cascade FWM-activated mode interactions are obtained. We find that the polarization of each filtered wavelength of PML tends to dithering/vibrating, which indicates that the well-defined SOP of conventional dissipative solitons is also broken in PML.

The fibre cavity is schematically shown in Fig. 2 (a). The SA is fabricated by filling reduced graphene oxide (rGO) flakes into cladding holes of a photonic crystal fibre (PCF): more details are in Supplementary 1. Since only a very small percentage ($10^{-7}$) of light passing through the PCF will interact with rGO, the thermal damage threshold can be increased substantially, a situation which is inaccessible for conventional ferule methods [30, 31]. Most importantly, the small nonlinear phase shift accumulated in each round trip makes it possible to observe the formation process of PML. The nonlinear transmission of the SA shown in Fig. 2 (b) indicates a modulation depth of ~ 24%. The dispersion of 1 m EDF, 19.5 m DCF, and 14.5 m SMF are 15.7, -38, and 18 ps/nm/km, respectively, so that the net normal dispersion of the cavity is 0.737 $ps^2$. The nonlinear response of SA and the dispersion variation are key in generating PI, which have been discussed in Supplementary 2.

Setting the power level of two pump lasers at 400 mW, and by rotating PC carefully, we observe a stable PML regime with easiness. The typical oscilloscope trace in Supplementary 3 contains no significant intensity fluctuations. The radio frequency spectrum exhibits a contrast ratio of 90 dB at the fundamental frequency of 5.82 MHz, suggesting that PML has very good stability. The autocorrelation trace in Fig. 2 (c) exhibits a pedestal with a full width at half maximum (FWHM) of 30.5 ps, and the coherent peak is about 416 fs. The FWHM of the optical spectrum detected by a conventional OSA in Fig. 2 (d) is 16 nm, and good agreement is found between 100 single-shot spectra detected by DFT together with time-averaged optical spectrum. The 6 consecutive single-shot spectra in Fig. 2 (e) indicate great fluctuations, which are invisible in the averaged optical spectrum. The details of the measurement are described in the Methods section.

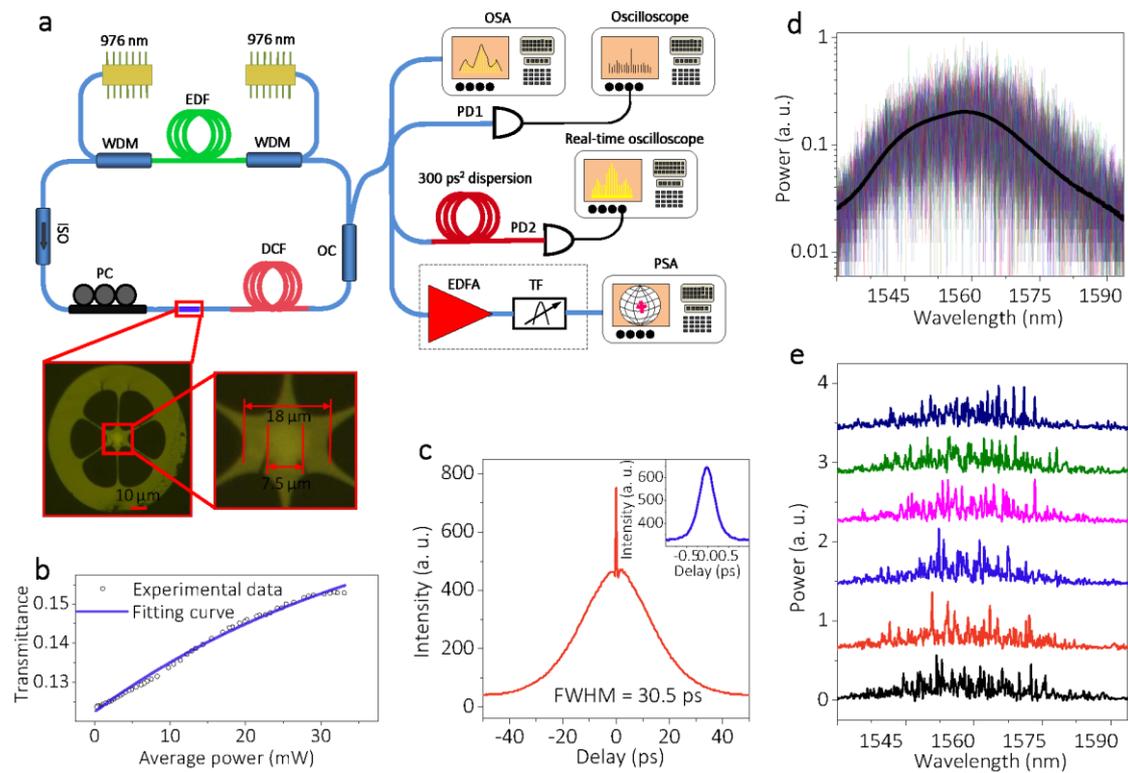

Figure 2. Laser setup and typical outputs for pump power at 800 mW. (a) EDF, erbium-doped fibre (Liekki ER 80-8/125); WDM, wavelength division multiplexer; ISO, polarization independent optical isolator; PC, polarization controller; DCF, dispersion compensation fibre; OC, optical coupler; OSA, optical spectrum analyzer; PD, photo-detector; EDFA, erbium-doped fibre amplifier; TF, tunable filter;

PSA, polarization state analyzer. The inset is the cross section of the PCF; more details are gven in the Supplementary. (b) Nonlinear transmission of the SA as a function of average power. (c) Autocorrelation trace: the inset is the coherent peak in a larger scale. (d) 100 single-shot spectra measured by DFT (colorized lines), and the black line is the averaged spectrum measured with an OSA. (e) 6 consecutive single-shot spectra showing enhanced spectral fluctuations in PML.

PML is obtained via rotating the PC, through which the nonlinear phase is detuned, and also the phase-matching condition of cascaded FWM is adjusted. Figure 3 depicts results under different PC states with a fixed pump power of 800 mW. The averaged optical spectra in Fig. 3 (a) suggest that the CW laser is destabilized by PI at first, then additional longitudinal modes are activated through cascaded FWM among different PI gain lobes, finally leading to PML when further optimizing the intracavity polarization evolution. The center mode and the four new sideband mode wavelengths, namely, $\lambda_{2AS}$, $\lambda_{1AS}$, $\lambda_{1S}$, $\lambda_{2AS}$, all satisfy the energy conservation of FWM. The PI sideband spacing is 0.438 THz, which is in relatively good accordance with the calculated value of Fig.1(b) for the estimated intracavity power level of about 5W of the central mode quasi-CW pulses.

The corresponding single-shot spectra contain more information about the PML evolution process. Primarily, a stable and highly coherent pulse train is generated for PC state 1: a much broader optical spectrum is formed in the single-shot spectrum. In this case, DFT filters automatically the giant CW mode component that does not participate to pulse formation. Considering that the resolution of the DFT is 0.2 nm (0.05 THz), PI gain lobes contained in this single-shot spectrum are reliable. We notice that those parametric gain lobes come entirely from the CW state. A further rotation of the PC (from PC state 1 to 3) primarily leads to secondary PI gain lobes. During this process, energy is transferred from the center mode to the parametric gain lobes at distant wavelengths, and a giant envelope is gradually imposed on the pulse train (Fig. 3 (f)) (More details are in the Supplementary 4). Continuously biasing the PC (from PC state 4 to 6) leads to supporting additional longitudinal modes, which eventually deplete the giant CW mode. The single-shot spectra appear as localized structures in two dimensions, as shown in Figs. 3 (g) and (h). The spliced spectra in Fig. 3 (e) for different round trips show that the well-defined structure in the averaged optical spectrum disappears in the single-shot spectrum. Finally, when the energy of the CW component is completely transferred into sidebands, the Q-switched-like envelope gradually diminishes, as more longitudinal modes are activated, until stable PML is formed.

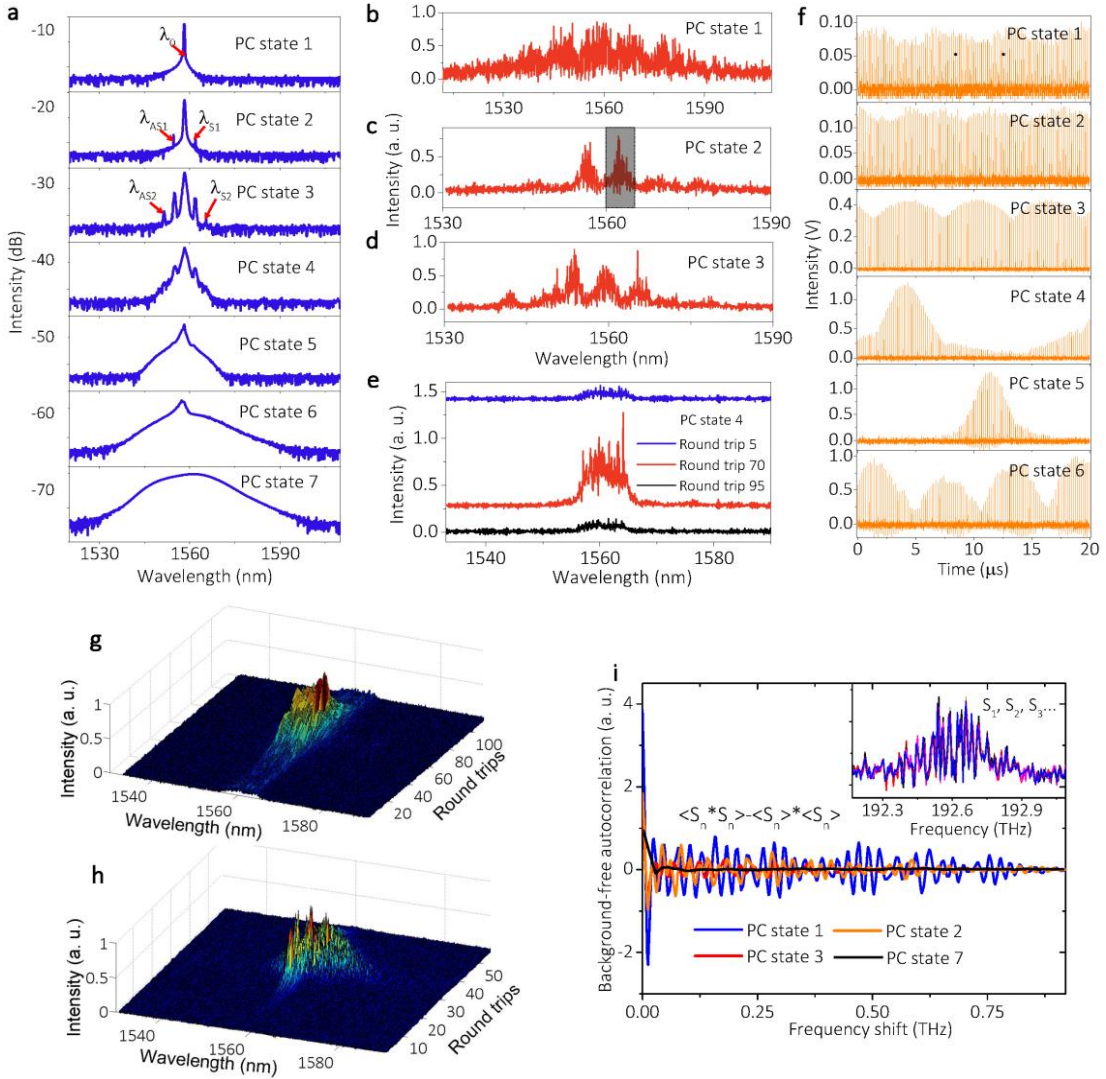

**Figure 3. Evolution of PML with polarization detuning.** (a) Averaged optical spectra for different PC states. The center wavelength $\lambda_0$ is 1558.43 nm, and the four new wavelengths are $\lambda_{2AS}$=1551.4 nm, $\lambda_{1AS}$=1554.8 nm, $\lambda_{1S}$=1561.6 nm, $\lambda_{2S}$=1565.1 nm. (b)-(e) Corresponding single-shot spectra for different PC states in (a). (f) Corresponding temporal pulse train detected by PD1. (g) & (h) Consecutive single-shot spectra under PC states 4 and 5. (i) Autocorrelation analysis of the first Stokes region in single-

shot spectra for different PC states. The inset contains 5 consecutive single-shot spectra within the first Stokes region (shaded region in Fig. 3 (c)) under PC state 2, where only small fluctuations are shown.

We investigated the coherence of the PI gain lobes based on a statistical analysis of the spectral correlations as described in Ref. 10. The background-free correlations of the first Stokes region (0.9 THz) for different PC states are shown in Fig. 3 (i), where a great difference is shown with respect to the case of MI in a conservative system [10]. Quasi-periodic fluctuations of the autocorrelation centered at zero indicate cooperative (positive value) and competitive (negative value) interactions of the individual cavity modes. Most importantly, the positive center peak with zero frequency shift, and the deep negative valley diminish gradually with polarization detuning. The same properties are observed in other Stokes and anti-Stokes regions. Yet, this correlation disappears when considering the whole spectrum (more details are given in Supplementary 4). When PML is formed, the autocorrelation trace of the first Stokes region finally evolves into to a flat zero-background, indicating that each mode is only correlated with itself. This loss of correlation is the representation of the loss of coherence.

Tuning of the pump strength may also form PML. Keeping the PC in state 7, we record the averaged optical spectra, temporal trace, and single-shot spectra for various levels of pump power. Figure 4 (a) reveals that more longitudinal modes are activated when the pump strength grows larger: correspondingly, irregular pulses with high intensities are formed as shown in Fig. 4 (b). We notice that temporal evolutions in Fig. 4 are highly different from what observed with PC detuning. The temporal symmetry of leading and trailing pulse edges, which is apparent in Fig. 3 (f), is absent when the pump power is varied. In this case the pulse intensity has a highly asymmetric temporal profile: it increases gently with an exponential trend in the leading pulse tail, while it decreases abruptly in the trailing edge.

The observed temporal asymmetry grows larger for stronger pumping levels. For low pump strengths, the pulse packets are localized between zero intensity regions (i.e., they are located in correspondence with the giant quasi-CW pulse). On the other hand, more packets appear when the pump power is high enough. For example, whenever the pump power is set at 750 mW, pulse packets merge with each other. Even higher pump powers flatten the pulse energy in each round-trip by coupling among different frequencies. Those properties clearly illustrate the onset of intracavity parametric frequency conversion, a nonlinear process that exhibits optical bistability when the phase matching is optimized. In fact, as shown by Fig.4(c), phase-locking of longitudinal modes leads to a hysteresis loop for the average output power as the pump power is varied in opposite directions. The periodicity of the pulse packets leads to the appearance of discrete frequency components in the average optical spectra of Fig. 4(a), due to the limited detection time of our OSA.

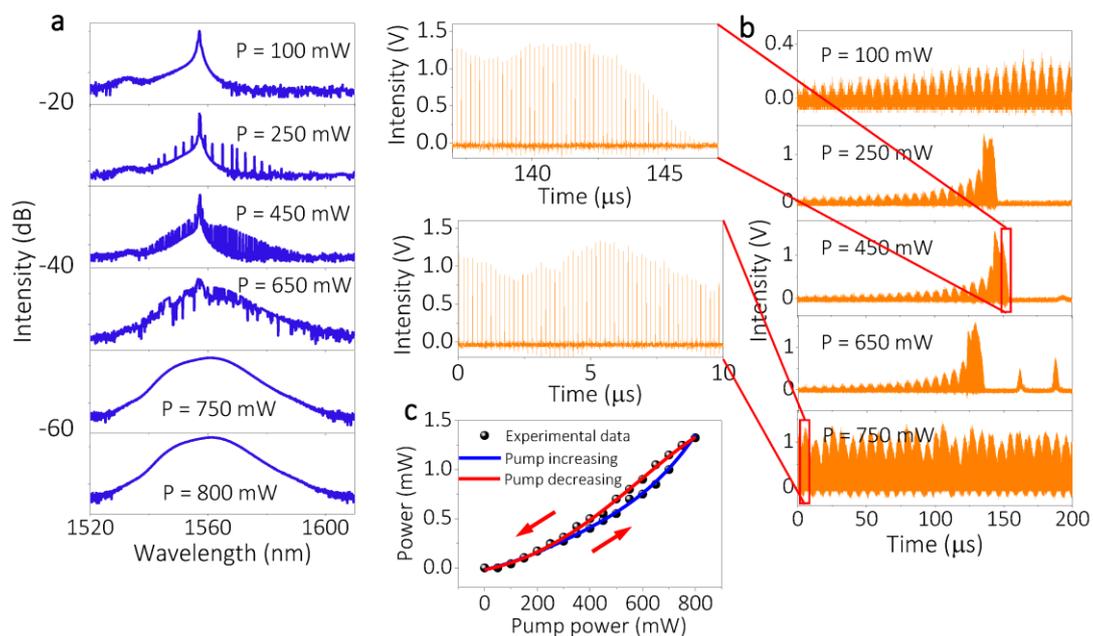

Figure 4. Evolution of PML as the pump strength is varied. (a) Averaged optical spectra for different pump power levels. (b) Corresponding temporal traces detected by PD1. The insets illustrate the cavity output over a smaller time range. (c) Hysteresis effect of average output power under different pumping regimes (increasing vs. decreasing pump power).

Although the average optical spectra in Fig. 4 (a) show that the giant quasi-CW mode remains present until the pump power is larger than 650 mW, the corresponding single-shot spectra reveal that the CW mode is depleted when the pump power is relatively small. Figure 5 (a) illustrates single-shot spectra for 100 mW pump power: as can be seen, quasi-CW and narrowband spectra are generated over the first few round trips, until new sidebands appear on both Stokes and anti-Stokes sides after further laser circulations. The narrow spectral region around the initial CW mode originates from self-phase modulation (SPM) induced spectral broadening. A depletion of this region is clearly shown in correspondence of the generation of distant sideband frequencies for round trip numbers larger than 40. Pump mode depletion deteriorates the QPM phase matching condition of PI, and leads to an abrupt switching to CW lasing. However, the CW transfers its energy quickly to sidebands for larger pump powers. This is accompanied by a sudden evolution from a narrow quasi-CW spectrum to a set of broad but randomly distributed comb lines, as shown in Fig. 5 (b). As the number of round trips that are necessary for accumulating a significant nonlinear phase shift decreases with increasing pump strength, the larger the pump power, the quicker the generation of pulse clusters. When the pump power is large enough, a broad comb with randomly spaced lines can be generated even within a single round trip.

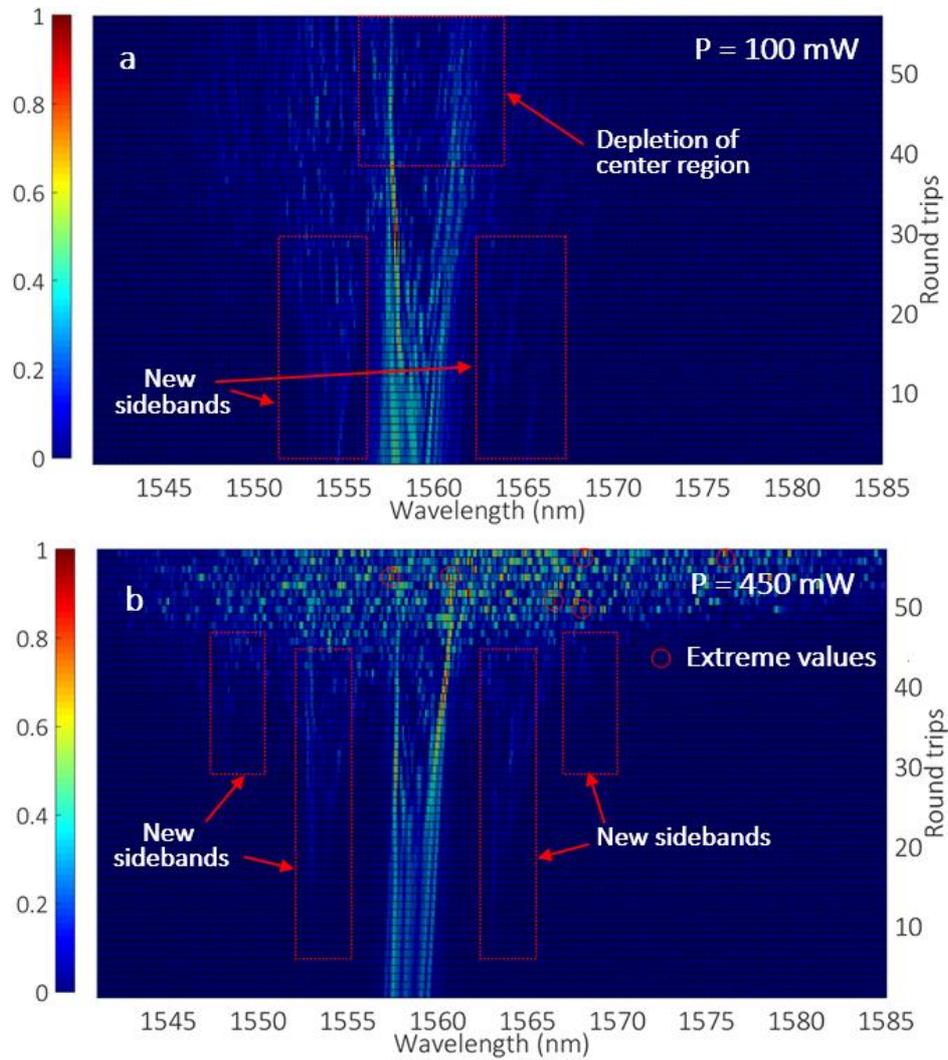

Figure 5. Single-shot spectra for pump powers of (a) 100 mW or (b) 450 mW, respectively. Regions doted by red circles are frequencies with extreme intensities, which could indicate the presence of spectral optical rogue waves.

In a PML based on the parametric frequency conversion process, different sets of longitudinal modes correspond to different portions of the pulse packets. To show this, we filtered out the PML output spectrum by a wavelength tunable filter with a bandwidth of 1 nm, and Fig. 6 (a) depicts the filtered spectra. The corresponding temporal signals are shown in Fig. 6 (b). It is clear that the center wavelength region (3), where SPM is the dominant effect, is responsible for the generation of the dense pulse train. For wavelengths far away from the center, the number of pulses in each packet decreases. Those results are totally different from the case of a highly coherent pulse train, such as a stable dissipative soliton laser, where the different frequencies contribute equally to the output pulses, and a filtering of output merely leads to broader pulse durations.

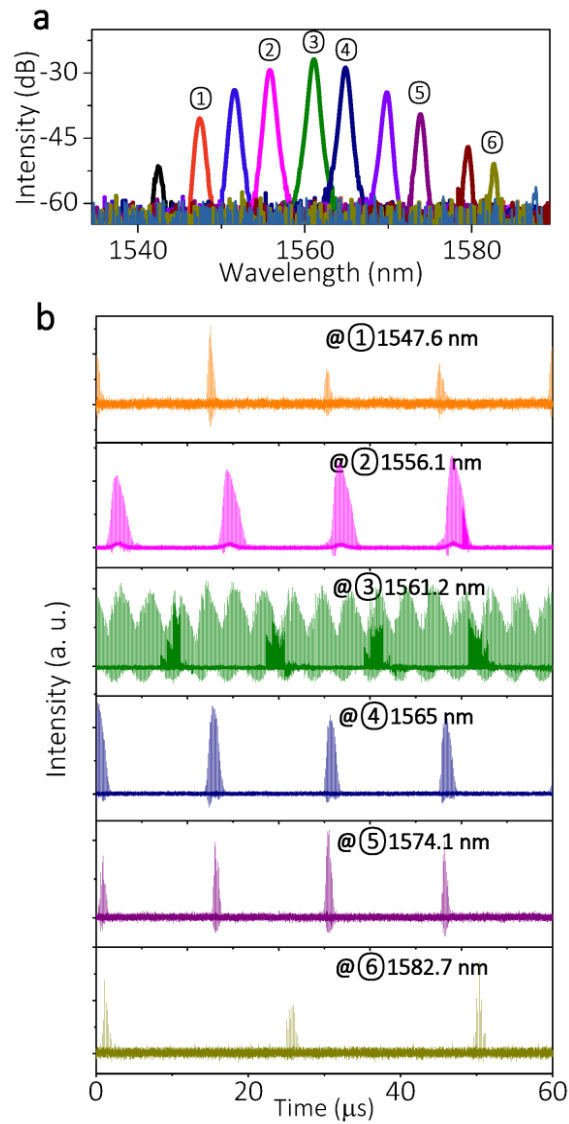

Figure 6. (a) Filter spectra with different center wavelengths, and (b) corresponding temporal traces. Index (3) indicates the central spectral region, while (2) and (4) indicate the primary Stokes and anti-Stokes gain lobes.

The loss of coherence is also reflected in the SOP of the filtered PML output. Figure 7 depicts the polarization states for different wavelengths and pump powers. Primarily, we observed that the SOP of the center wavelength remains a fixed point on the Poincaré sphere. On the other hand, Fig.7 shows that wavelengths far away from the center mode bifurcate into a cross-like configuration of output SOPs. The two perpendicular lines in each cross are aligned with either a meridian or a parallel curve on the Poincaré sphere, respectively, which indicates that two separate processes are present that lead the azimuth and ellipticity angles of the output SOPs to evolve separately. The four directions of SOP evolution originate from vector FWM processes. Uneven fluctuations of the PI sidebands are responsible for stretching the corresponding SOP in and out of the cross. Besides, the two SOP curves along the parallel of the Poincaré sphere tend to bifurcate when a red-shifted wavelength component appears, such as in wavelength regions 5 & 6 for the pump power of 250 mW. This bifurcation of polarization may be determined by the coexistence of PMI-like process and XPMI-like process.

Furthermore, irregular polarization states located outside of the main polarization directions emerge for red-shifted sidebands whenever the pump power reaches 250 mW: SOP scattering aggravates over the whole wavelength span for the pump power of 450 mW. We attribute the polarization deviation from the cross shape to chaotic competition of cascaded FWM processes, including both scalar and vector FWMs, through which the energy of longitudinal modes with a SOP along the cross is transferred into newly-generated SOPs. When the pump power is larger than 600 mW, additional longitudinal modes are populated, and the SOPs of filtered wavelengths appear as totally random, owing to enhanced spectral energy transfer processes. This randomization of output SOP is associated with the loss of temporal coherence.

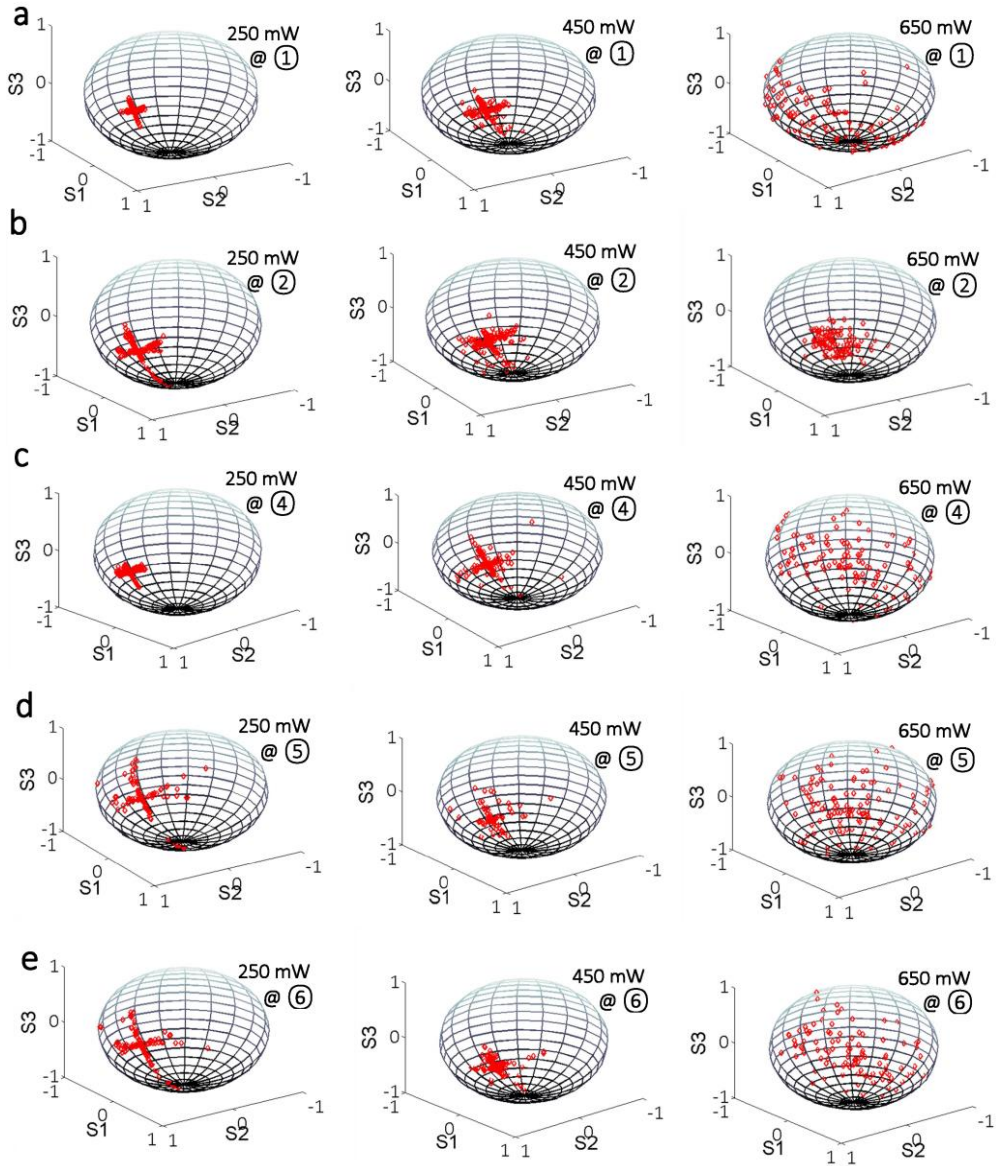

Figure 7. Experimentally measured polarization states for filtered wavelengths under various pump powers The polarization states of center wavelength at (3) are always as a fixed point in the Poincaré sphere for different pump powers (not shown).

When considering the complex SOP evolutions in a PML, we envisage that such a light source can be utilized for highly coherent quantum entanglement applications. The all-fibre configuration makes such a source compatible with low loss optical networks. Moreover, frequencies with extreme intensities can be found at unpredictable positions on the broad spectrum. In analogy with incoherent spectral optical solitons [32], our experiments indicate that that the loss of coherence in a PML could lead to the generation of incoherent spectral optical rogue waves. Such analysis will be the subject of subsequent detailed investigations. Given the analogies between the dynamics of PML and Kerr-combs, the analysis of spectra based on the DFT in our experiment could be helpful in the understanding the dissipative temporal and spectral pattern formation in microresonators, where is extremely hard to perform DFT detection due to the large bandwidth of the spectral output.

In summary, we experimentally investigated the dynamics of PML formation based on parametric instability followed by cascade FWM, controlled by polarization detuning. The PML scatters first its energy from a center wavelength mode to distant cavity modes via highly coherent parametric instability. Subsequently, all longitudinal modes are populated via cascaded FWM among the different orders of the parametric gain lobes. We found that mode interactions seeded by the highly coherent PI can either be competitive or cooperative, until eventually the laser system fully looses its coherence. In the incoherent stage, energy coupling among the various modes occurs when adjusting the intracavity SOP and when increasing the external pumping strength, until a bifurcation into random SOPs is identified. The observed dynamics of spatio-temporal evolutions in a PML, leading to the loss of coherence, are beneficial for advancing understanding of supercontinuum and Kerr-combs phenomena. These results may also provide a new perspective for the understanding of the loss of coherence in various nonlinear wave processes, even beyond the domain of nonlinear optics.

## Methods

**Measurement methods.** The laser output is monitored by two kinds of detectors (PD1, 350 MHz; PD2, 50 GHz). The bandwidth of the oscilloscope is 1 GHz. As the rise time of PD1 is much larger than the duration of PML, the recorded pulse amplitude is proportional to the pulse energy. An autocorrelator with a delay resolution of 6 fs and an optical spectrum analyzer with a resolution of 0.02 nm are utilized. The SOP is measured by a high speed polarization measurement system. The real time oscilloscope in the DFT has a bandwidth of 20 GHz, and the signals are stretched by dispersive fibre with the dispersion of 300 ps², and subsequently fed to a 50 GHz PD2 for frequency-to-time transformation. To ensure the detector is not saturated, a variable optical attenuator is inserted between the detector and the laser output.

**Numerical simulation.** The parametric gain of Fig. 1 (b) is computed from the Floquet method applied to the nonlinear Schrödinger equation (NLSE) with varying dispersion and nonlinear coefficients [33]. The evolution of the optical field $\psi$ in fibre laser can be described by the NLSE that includes Kerr nonlinearity $\gamma$ and second-order dispersion $\beta_2$

$$i\frac{\partial \psi}{\partial z} - \frac{\beta_2}{2}\frac{\partial^2 \psi}{\partial t^2} + \gamma |\psi|^2 \psi + i\frac{\alpha}{2}\psi = 0 \tag{1}$$

The influence of linear losses is included through the coefficient $\alpha$ (negative and positive values leading to distributed amplification and loss, respectively). In Eq. (1) we did not include higher-order dispersion or Raman scattering. These effects do not have a noticeable influence on the MI spectral dynamics in the fibre laser. With the change of variables $U=\psi\exp(\alpha z/2)$, one obtains from (1)

$$i\frac{\partial U}{\partial z} - \frac{\beta_2}{2}\frac{\partial^2 U}{\partial t^2} + \gamma'(z)|U|^2 U = 0 \tag{2}$$

where $\gamma'=\gamma\exp(-\alpha z)$. In the simulations, we consider a Kerr nonlinearity of $\gamma=2$ W⁻¹km⁻¹, a spatial period of the dispersion variation equal to the cavity length (35 m), comprising 19.5 m span of DCF (with dispersion $D$ = -38 ps/nm/km and loss of 0.5 dB/km), 14.5 m span of SMF (with dispersion $D$ = 18 ps/nm/km and loss of 0.2 dB/km), and 1 m EDF (with dispersion $D$ = 15.7 ps/nm/km and a gain that compensates cavity loss exactly). Before the EDF, we inserted a 1dB of lumped loss to take into account out-coupling loss and connector losses. The pulse duration of PC state 1 is about 0.04 ns. According to the relative pulse energy value detected by low speed PD1, the average power of the pulses in the cavity is about 1.2 mW. Thus, the peak power is about 5.15 W. We may write the perturbed CW solution of the NLSE (2) as

$$U(z,t) = \left[\sqrt{P} + u(z,t)\right]\exp\{iPz\} \qquad (3)$$

where we suppose $|u|^2 \ll P$, so that one obtains the linearized equation for $u(Z,T)$

$$iu_Z = b^2 \frac{d(Z)}{2} u_{TT} + g'(Z)\left(u + u^*\right) \qquad (4)$$

where we define $b^2 = L_{nl}/L_d$, (the nonlinear length $L_{nl}=1/(\gamma P)$, the dispersion length $L_d=t_0^2/|\beta_{2ave}|$, and the reference time unit $t_0=1$ ps). Therefore, $Z=z/L_{nl}$, and $T=t/t_0$. By writing the solution of (4) as the sum of Stokes and anti-Stokes sidebands, $u(Z,T)=a(Z)\exp\{i\Omega T\}+b^*(Z)\exp\{-i\Omega T\}$, we obtain two coupled linear ordinary differential equations (ODEs) with periodic coefficients for $a(Z)$ and $b(Z)$

$$a_Z = i\frac{b^2}{2}W^2 d(Z)a + ig'(Z)(a+b) \qquad (5a)$$

$$b_Z = -i\frac{b^2}{2}W^2 d(Z)b - ig'(Z)(a+b) \qquad (5b)$$

Equations (5) are equivalent to a single, second-order ODE, and a linear stability analysis can be rigorously carried out numerically by the Floquet theory, which is analogous to Bloch wave theory in solid state physics. By defining the solution vector of (5) as $s=(a,b)$, and choosing the two fundamental initial conditions $s_1(Z=0)=(1,0)$ and $s_2(Z=0)=(0,1)$, we obtains the principal solution matrix $S=[s_1(L')^t, s_2(L')^t]$ (where $t$ denotes the vector or matrix transpose) from the solution of (5) at $Z=L'$ (where $L'=L/L_{nl}$). According to Floquet's theorem, the eigenvalues of S, or Floquet's multipliers $\lambda=\exp(\eta_F+i\sigma)$, such that $|\lambda|>1$ yields the linear instability of the CW with respect to the growth of sidebands with frequency detuning $\Omega$. Since the scattering matrix after an integer number n of periods is simply $S^n$, the Floquet's multipliers to the MI gain G after one period of the dispersion and power oscillation in the laser cavity can be related by $G=2\eta_F/L'$.

**References**

1. Turitsyna, E. G. *et al*. The laminar–turbulent transition in a fibre laser. *Nat. Photon*. **7**, 783-786 (2013).

2. Dudley, J. M., Dias, F., Erkintalo, M. & Genty, G. Instabilities, breathers and rogue waves in optics. *Nat. Photon*. **8**, 755–764 (2014).

3. Elizabeth, A. D., Claussen, N. R., Thompson, S. T. & Wieman, C. E. Atom–molecule coherence in a Bose–Einstein condensate. *Nature* **417**, 529-533 (2002).

4. Hanson, R. & Awschalom, D. D. Coherent manipulation of single spins in semiconductors. *Nature* **453**, 1043-1049 (2008).

5. McNeil, B.W. J. & Thompson, N. R. X-ray free-electron lasers. *Nat. Photon*. **4**, 814–821 (2010).

6. Avila, K. *et al*. The onset of turbulence in pipe flow. *Science* **333**, 192–196 (2011).

7. Fermann, M. E. & Hartl, I.  Ultrafast fibre lasers. *Nat. Photon*. **7**, 868-874 (2013).

8. Grelu, P. & Akhmediev, N. Dissipative solitons for mode-locked lasers. *Nat. Photon.* **6**, 84–92 (2012).

9. Oktem, B., Ülgüdür, C. & Ilday, F. Ö. Soliton–similariton fibre laser. *Nat. Photon*. **4**, 307-311 (2010).

10. Solli, D. R., Herink, G., Jalali, B. & Ropers, C. Fluctuations and correlations in modulation instability. *Nat. Photon.* **6**, 463–468 (2012).

11. Gao, L., Zhu, T., Liu, M. & Huang, W. Cross-phase modulation instability in mode-locked laser based on reduced graphene oxide. *IEEE Photon. Technol. Lett.* **27**, 38-41 (2015).

12. Smith, N. J. & Doran, N. J. Modulational instabilities in fibers with periodic dispersion management. *Opt. Lett*. **21**, 570-572 (1996).

13. Kudlinski, A. *et al*. Simultaneous scalar and cross-phase modulation instabilities in highly birefringent photonic crystal fiber. *Opt. Express* **21**, 8437-8443 (2013).

14. Horowitz, M., Barad, Y. & Silberberg Y. Noiselike pulses with a broadband spectrum generated from an erbium-doped fiber laser. *Opt. Lett*. **22**, 799-801 (1997).

15. Zhao, L. M., Tang, D. Y., Cheng, T. H., Tam, H. Y. & Lu, C. 120 nm bandwidth noise-like pulse generation in an erbium-doped fiber laser. *Opt. Commun*. **281**, 157-161 (2008).

16. Jeong, Y., Vazquez-Zuniga, L. A., Lee, S. & Kwon, Y. On the formation of noise-like pulses in fiber ring cavity configurations. *Opt. Fiber Technol*. **20**, 575-592 (2014).

17. Runge, A., Aguergaray, C., Broderick, N. & Erkintalo, M. Coherence and shot-to-shot spectral fluctuations in noise-like ultrafast fiber lasers. *Opt. Lett*. **38**, 4327-4330 (2013).

18. North, T. & Rochette, M. Raman-induced noiselike pulses in a highly nonlinear and dispersive all-fiber ring laser. *Opt. Lett*. **38**, 890-892 (2013).

19. Churkin, D. V. *et al*. Stochasticity, periodicity and localized light structures in partially mode-locked fibre lasers, *Nat. Commun*. **6**, 7004 (2015).

20. Tang, D. Y., Zhao, L. M. & Zhao, B. Soliton collapse and bunched noise-like pulse generation in a passively mode-locked fiber ring laser. *Opt. Express* **13**, 2289-2294 (2005).

21. Wang, Q. *et al.* All-fiber ultrafast thulium-doped fiber ring laser with dissipative soliton and noise-like output in normal dispersion by single-wall carbon nanotubes. *Appl. Phys. Lett.* **103**, 011103 (2013).

22. Peccianti, M. *et al.* Demonstration of a stable ultrafast laser based on a nonlinear microcavity. *Nat. Commun.* **3**, 765 (2012).

23. Herr, T. *et al.* Universal formation dynamics and noise of Kerr-frequency combs in microresonators. *Nat. Photon.* **6**, 480-487 (2012).

24. Ferdous, F. *et al.* Spectral line-by-line pulse shaping of on-chip microresonator frequency combs. *Nat. Photon.* **5**, 770-776 (2011).

25. Schröder, J., Vo, T. D. & Eggleton, B. J. Repetition-rate-selective, wavelength-tunable mode-locked laser at up to 640 GHz. *Opt. Lett*. **34**, 3902-3904 (2009).

26. Kudlinski, A. *et al.* Simultaneous scalar and cross-phase modulation instabilities in highly birefringent photonic crystal fiber. *Opt. Express* **21**, 8437-8443 (2013).

27. Smith, N. J. & Doran, N. J. Modulational instabilities in fibers with periodic dispersion management. *Opt. Lett*. **21**, 570-572 (1996).

28. Droques, M., Kudlinski, A., Bouwmans, G., Martinelli, G. & Mussot, A. Experimental demonstration of modulation instability in an optical fiber with a periodic dispersion landscape. *Opt. Lett.* **37**, 4832-4834 (2012).

29. Finot, C., Fatome, J., Sysoliatin, A., Kosolapov, A. & Wabnitz, S. Competing four-wave mixing processes in dispersion oscillating telecom fiber. *Opt. Lett*. **38**, 5361-5364 (2013).

30. Bao, Q. L. *et al*. Atomic-layer graphene as a saturable absorber for ultrafast pulsed lasers. *Adv. Funct. Mater*. **19**, 3077-3083 (2009).

31. Sun, Z. P. *et al*. Graphene mode-locked ultrafast laser. *ACS. NANO*. **4**, 803-810 (2010).

32. Picozzi, A., Pitois, S. & Millot, G. Spectral incoherent solitons: a localized soliton behavior in the frequency domain. *Phys. Rev. Lett*. **101**, 093901 (2008).

33. Finot, C., Feng, F., Chembo, Y. & Wabnitz, S. Gain sideband splitting in dispersion oscillating fibers. *Opt. Fiber Tech*. **20**, 513-519 (2014).

### Acknowledgments

This work was supported by the Natural Science Foundation of China (No. 61405020, 61475029, and 61377066) and the Science Fund for Distinguished Young Scholars of Chongqing (No. CSTC2014JCYJJQ40002). The work of Stefan Wabnitz was supported by the Italian Ministry of University and Research (MIUR) (2012BFNWZ2).

## Author contribution

Lei Gao proposed the laser system, performed the main experiment, and wrote the main manuscript text. Tao Zhu improved the designs of the system, instructed the experiments, and supervised the whole project. Stefan Wabnitz developed the theory of parametric instability and performed corresponding gain simulations. Min Liu, and Wei Huang contributed to the scientific discussion and improved the manuscript presentation. All authors discussed the results and substantially contributed to the manuscript.

## Additional information.

Supplementary information is available in the online version of the paper.

The authors declare no competing financial interests.

## Figure captions

**Figure 1. PML dynamics based on stochastic parametric conversion. (a)** Primary gain sidebands, $\omega_{1S}$ and $\omega_{1AS}$, are generated from parametric instability: high-order gain lobes are activated with the optimization of the polarization state. Δ would be preserved in all sidebands due to energy conservation. Cascaded FWM populates longitudinal modes between the PI gain lobes. The insets are phase-matching processes of two different vector MI processes. The solid and dotted lines represent two orthogonal polarization components, respectively. In the PMI process, the polarizations of two sidebands are orthogonal to that of the pump, and are parallel to each other. In XPMI, sidebands are polarized at 45° from the pump laser polarization; **(b)** Numerically computed PI gain sidebands in the laser cavity with periodic dispersion and loss. The calculated sideband shift for the peak power of 5.15 W is about 490 GHz, which is close to the experimental value of 438 GHz. The discrepancy is mainly induced by the estimate error of the pulse average power.

**Figure 2. Laser setup and typical outputs for pump power at 800 mW.** (a) EDF, erbium-doped fiber (Liekki ER 80-8/125); WDM, wavelength division multiplexer; ISO, polarization independent optical isolator; PC, polarization controller; DCF, dispersion compensation fiber; OC, optical coupler; OSA,

optical spectrum analyzer; PD, photo-detector; EDFA, erbium-doped fiber amplifier; TF, tunable filter; PSA, polarization state analyzer. The inset is the cross section of the PCF; more details are gven in the Supplementary. (b) Nonlinear transmission of the SA as a function of average power. (c) Autocorrelation trace: the inset is the coherent peak in a larger scale. (d) 100 single-shot spectra measured by DFT (colorized lines), and the black line is the averaged spectrum measured with an OSA. (e) 6 consecutive single-shot spectra showing enhanced spectral fluctuations in PML.

**Figure 3. Evolution of PML with polarization detuning.** (a) Averaged optical spectra for different PC states. The center wavelength $\lambda_0$ is 1558.43 nm, and the four new wavelengths are $\lambda_{2AS}$=1551.4 nm, $\lambda_{1AS}$=1554.8 nm, $\lambda_{1S}$=1561.6 nm, $\lambda_{2S}$=1565.1 nm. (b)-(e) Corresponding single-shot spectra for different PC states in (a). (f) Corresponding temporal pulse train detected by PD1. (g) & (h) Consecutive single-shot spectra under PC states 4 and 5. (i) Autocorrelation analysis of the first Stokes region in single-shot spectra for different PC states. The inset contains 5 consecutive single-shot spectra within the first Stokes region (shaded region in Fig. 3 (c)) under PC state 2, where only small fluctuations are shown.

**Figure 4. Evolution of PML as the pump strength is varied.** (a) Averaged optical spectra for different pump power levels. (b) Corresponding temporal traces detected by PD1. The insets illustrate the cavity output over a smaller time range. (c) Hysteresis effect of average output power under different pumping regimes (increasing vs. decreasing pump power).

**Figure 5. Single-shot spectra for pump powers of (a) 100 mW or (b) 450 mW, respectively.** Regions doted by red circles are frequencies with extreme intensities, which could indicate the presence of spectral optical rogue waves.

**Figure 6. (a) Filter spectra with different center wavelengths, and (d) corresponding temporal traces.** Index (3) indicates the central spectral region, while (2) and (4) indicate the primary Stokes and anti-Stokes gain lobes.

**Figure 7. Experimentally measured polarization states for filtered wavelengths under various pump powers** The polarization states of center wavelength at (3) are always as a fixed point in the Poincaré sphere for different pump powers (not shown).